\documentclass[%
 reprint,
 amsmath,amssymb,
 aps,
 prb,
]{revtex4-1}

\usepackage{graphicx}
\usepackage{dcolumn}
\usepackage{bm}
\usepackage{epstopdf}

\graphicspath{{fig/}}

\begin{document}

\preprint{APS/123-QED}

\title{Spin-orbit coupling and magnetic interactions in Si(111):\{C,Si,Sn,Pb\} }
\author{D.I. Badrtdinov$^{1}$, S.A. Nikolaev$^{1}$, M.I. Katsnelson$^{1,2}$ and V.V. Mazurenko$^{1}$}

\affiliation{$^{1}$Theoretical Physics and Applied Mathematics Department, Ural Federal University, 620002 Ekaterinburg, Russia \\
$^{2}$ Institute for Molecules and Materials, Radboud University, Heyendaalseweg 135, 6525 AJ Nijmegen, The Netherlands
}

\date{\today}

\begin{abstract}
We study the magnetic properties of the adatom systems on a semiconductor surface Si(111):\{C,Si,Sn,Pb\} - ($\sqrt{3} \times \sqrt{3}$). On the basis of all-electron density functional theory calculations we construct effective low-energy models taking into account spin-orbit coupling and electronic correlations. In the ground state the surface nanostructures are found to be insulators with the non-collinear 120$^{\circ}$ N\'eel (for C, Si, Sn monolayer coverages) and 120$^{\circ}$ row-wise (for Pb adatom) antiferromagnetic orderings. The corresponding spin Hamiltonians with anisotropic exchange interactions are derived by means of the superexchange theory and the calculated Dzyaloshinskii-Moriya interactions are revealed to be very strong and compatible with the isotropic exchange couplings in the systems with Sn and Pb adatoms. To simulate the excited magnetic states we solve the constructed spin models by means of the Monte Carlo method. At low temperatures and zero magnetic field we observe complex spin spiral patterns in Sn/Si(111) and Pb/Si(111). On this basis the formation of antiferromagnetic skyrmion lattice states in adatom $sp$ electron systems in strong magnetic fields is discussed. 
\end{abstract}

\maketitle


\section{\label{sec:level1}Introduction}

\begin{figure*}[t]
\includegraphics[width= 0.9\textwidth]{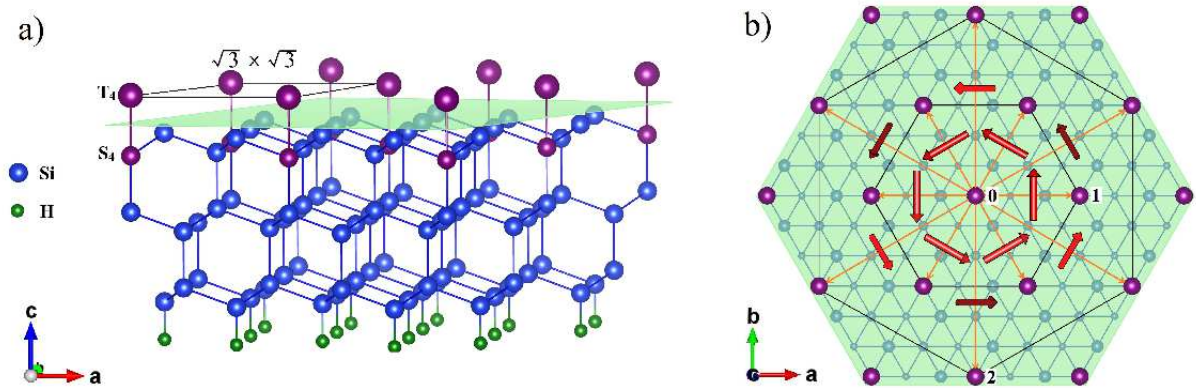}
\caption{a) Crystal structure of Si(111):X. Violet spheres denote T$_4$-S$_4$ positions of adatoms X, blue and green spheres show the silicon and hydrogen atoms, respectively. b) $ab$-plane. The interaction paths are shown with orange lines. Red arrows stand for the direction of DMI. For next nearest  neighbor interactions dark and light red arrows denote the Dzyaloshinskii-Moriya vectors with the negative and positive $z$ component, respectively. Crystal structures are visualized by using the VESTA software \cite{VESTA}.}
\label{im:Crystal}
\end{figure*} 

There is a special focus on the adatom systems Si(111):\{C,Si,Sn,Pb\}- ($\sqrt{3} \times \sqrt{3}$) formed by a silicon  surface (111) with the 1/3 monolayer coverage by C, Si, Sn or Pb adatoms. Being the physical realizations of the one-band Hubbard model on a triangular lattice, this family of the adatom materials demonstrates a remarkable variety of interesting physical properties. For instance, the scanning tunneling spectroscopy and photoemission spectroscopy experiments\cite{Modesti} on Sn/Si(111) demonstrated isostructural metal-insulator transition at $\sim 60$ K predicted by Profeta and Tosatti\cite{Tosatti} on the basis of LSDA+$U$ calculations. Another important phenomenon observed in the scanning tunneling microscopy experiments is a charge density wave state related to the redistribution of the valence electrons in the system.\cite{Carpinelli, Slezak}

On the theoretical side, the main efforts were concentrated on the construction and solution of minimal electronic models taking into account local and non-local Coulomb interactions.\cite{Hansmann, Hansmann_1, Lechermann, Li,Li1,Tosatti} These studies successfully reproduced experimentally observed metal-insulator transitions and charge ordering phase diagrams. Moreover, numerical simulations of the many-body Hamiltonians helped to resolve the existing discrepancies in different experiments suggesting different ordering phenomena.\cite{Hansmann_2}

Much less attention has been paid to the magnetic properties of the Si(111):\{C,Si,Sn,Pb\} systems. At the moment there is no consistent description of the magnetic ground state as well as of the excited states at finite temperatures and magnetic fields. For instance, first-principles simulations\cite{Lechermann} of the adatom system with Sn revealed that the 120$^{\circ}$ antiferromagnetic state has the lowest total energy, although it was shown that the magnetic moments are strongly delocalized. In turn, the authors of Ref. \onlinecite{Li} reported on the formation of the so-called collinear row-wise magnetic ordering in the Sn/Si(111) system formed due to hopping processes beyond nearest neighbors. Such a magnetic model was motivated by the comparison of the angle-resolved photoemission spectroscopy (ARPES) experiment and dynamical cluster approximation spectra. However, there is still no direct experimental confirmation of the row-wise magnetic state. Last but not least, spin-orbit coupling was not taken into account in these studies. However, it can be very important and lead to nontrivial topological properties as it follows from the DFT results for Si/Si(111) presented in Ref.\onlinecite{Fu}. The situation with spin-orbit coupling can be even more interesting in the case of heavy adatoms such as Sn and Pb.

In this paper we perform a comprehensive theoretical description of magnetic properties in the Si(111): \{C,Si,Sn,Pb\} systems in the ground state as well as at finite temperatures and magnetic fields. Our approach combines first-principles simulations within density functional theory, construction of the low-energy models taking into account spin-orbit coupling and electronic correlations in the Wannier function basis and determination of the magnetic interactions by means of the superexchange theory. The zero-temperature Hartree-Fock solution of the constructed electronic models reveals non-collinear 120$^{\circ}$ N\'eel states for the systems with carbon, silicon and tin, while the 120$^{\circ}$ row-wise state is found to be more stable in Si(111):Pb. Here, we argue that the formation of a magnetic order in the Si(111):\{X\} systems is a joint effect of spin-orbit coupling, direct exchange interactions between neighboring Wannier functions and hopping parameters beyond nearest neighbors.

Another important result we obtained by means of classical Monte Carlo simulations is the complex spin patterns, such as interpenetrating spin spirals stabilized in the Si(111):\{Sn, Pb\} systems at low temperatures. These non-trivial structures are formed due to strong Dzyaloshinskii-Moriya interactions (DMI) between nearest neighbors on the triangular lattice and, in principle, can be experimentally observed by using spin-polarized scanning tunneling microscopy.\cite{Wiesendanger, Wiesendanger1} Finally, at extremely large magnetic fields ($\sim$ 200 T) we predict the stabilization of an antiferromagnetic skyrmion lattice state.

\section{\label{sec:level1}Results of DFT+SO calculations}
To simulate electronic and magnetic properties of the Si(111):\{C,Si,Sn,Pb\} systems, we have performed first-principles calculations within density functional theory (DFT)\cite{DFT} using the generalized gradient approximation (GGA) with the Perdew-Burke-Ernzerhof (PBE) exchange-correlation functional.\cite{PBE} To this end, we have employed Quantum Espresso\cite{espresso} and Vienna ab-initio simulation package (VASP).\cite{Kresse, Furthmuller} In these calculations, we set an energy cutoff in the plane-wave decomposition to 400 eV and the energy convergence criteria to 10$^{-4}$ eV. For the Brillouin-zone integration a 20$\times$20$\times$1 Monkhorst-Pack mesh was used.

The simulated atomic structures of the Si(111):\{C,Si,Sn,Pb\} systems are presented in Fig.\ref{im:Crystal} and contain three layers of silicon, one monolayer of adatoms and a hydrogen slab, as described in Ref.\onlinecite{Hansmann}. Here, adatoms occupy the T$_4$ positions in Si/Si(111), Sn/Si(111) and Pb/Si(111),\cite{Hansmann_1,Lechermann} while in the case of the C/Si(111) adatoms are in the S$_4$ underlayer positions.\cite{Pignedoli}
The optimized atomic structures are consistent with those reported in previous studies.\cite{Hansmann_1, Lechermann, Pignedoli}.

\begin{figure}[!h]
\includegraphics[width=0.49\textwidth]{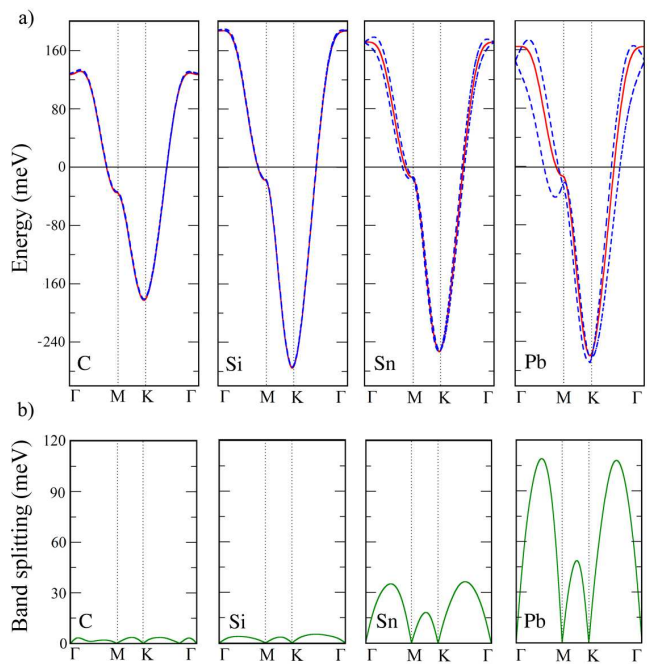}
\caption{a) Band structures of Si(111):\{C,Si,Sn,Pb\} near the Fermi level as obtained from DFT (red solid line) and DFT+SO (blue dashed line) calculations. b) Band splitting (in meV) due to spin-orbit coupling.}
\label{im:BANDS_SO}
\end{figure}  

Band structures calculated within DFT demonstrate the main peculiarity of the systems that is one well-separated doubly-degenerate band located near the Fermi level, which is further split when spin-orbit coupling (DFT+SO) is taken into account (Fig.\ref{im:BANDS_SO}b). This splitting strongly depends on the adatom type and varies from 3.5 meV (for carbon) to 110 meV (in the case of Pb adatoms). Thus, within this family of the surface nanostructures one can probe weak and strong limits of spin-orbit coupling in a strongly correlated material. 

\section{Wannier functions}
\begin{figure}[t]
\includegraphics[width=0.49\textwidth]{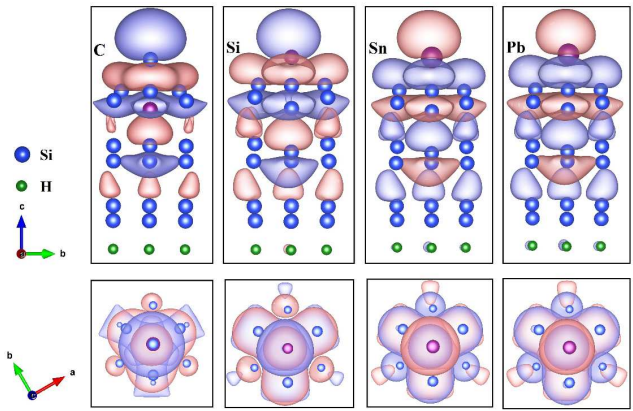}
\caption{Maximally localized Wannier functions describing the band at the Fermi level in Si(111):\{C, Si, Sn, Pb\}. Violet spheres denote adatoms (center of the Wannier function).} 
\label{im:Wannier}
\end{figure}

To parametrize the DFT+SO spectra and  construct the corresponding low-energy models we used maximally localized Wannier functions.\cite{wannier90,wannier901,wannier902} As it is shown in Fig. \ref{im:Wannier}, being centered at the adatom $pz$-orbitals the resulting Wannier functions are strongly delocalized (Fig.\ref{im:Wannier}). Their spread of the Wannier functions in Si(111):\{C,Si,Sn,Pb\} (Table \ref{tab:Bare_parameters}) is much larger than that one observes in $3d$ transition metal compounds with strong hybridization effects. For instance, the WF spread in a copper oxide \cite{Danis} is about 4.5 \AA$^{2}$. As we will show below, such a delocalization of the magnetic orbitals leads to an additional ferromagnetic contribution to the total exchange interaction between nearest neighbours in the system.

Another sign of the magnetic orbital delocalization is the contribution of the atomic-like $pz$ orbital of the adatom to the Wannier function. From Table \ref{tab:Bare_parameters} one can see that the $pz$ orbital contributes about 28 \% to the electronic density around the Fermi level. This value is two times smaller than that calculated for low-dimensional cuprate \cite{Danis}.

\section{Orbital magnetization}
According to our DFT results, Si(111):\{C,Si,Sn,Pb\} surface nanostructures are characterized by strong spin-orbit coupling.  The $pz$ atomic orbital of the adatom (head of the magnetic orbital) corresponds to L=0 and, as the result, gives zero contribution orbital magnetization.  However, in the situation of the strong delocalization of the Wannier function one can expect that there could be a non-zero net orbital magnetization as described in Ref.\onlinecite{Vanderbilt,Nikolaev_orb}. To estimate it we performed calculations by using procedure realized in the Wannier 90 package. In these calculations we use minimal ($\sqrt{3} \times \sqrt{3}$) unit cell with ferromagnetic configuration. 

Due to the strong hybridization and spin-orbit coupling, the resulting spin moment of the unit cell is considerably suppressed in the case of the Sn and Pb adatoms (Table \ref{tab:Bare_parameters}). The calculated total magnetic moment in Sn/Si(111) system agrees with results of Ref.\onlinecite{Lechermann}. 

It was found that the orbital magnetization is close to zero ($\sim$ 10$^{-3}$   $\mu_{B}$) for all the systems in question. Thus we conclude that the g-factor is purely spin one, $g$ = 2. This result will be used in Section VIII for estimating critical magnetic fields of skyrmion formation.  

\section{Low-energy model}
To describe electronic and magnetic properties of the adatom systems we use an effective electronic model taking into account spin-orbit coupling and electronic correlations in the Wannier function basis: 
\begin{eqnarray}
\hat{\cal H}=\sum_{i j,\sigma\sigma'}t_{ij}^{\sigma\sigma'}\hat{a}_{i \sigma}^{+}\hat{a}_{j \sigma'}+\frac{1}{2}\sum_{i ,\sigma\sigma'}U \,\hat{a}_{i \sigma}^{+}\hat{a}_{i \sigma'}^{+}\hat{a}^{}_{i \sigma'}\hat{a}^{}_{i \sigma} \nonumber \\ 
+ \frac{1}{2}\sum_{i j,\sigma\sigma'} V_{ij}\,\hat{a}_{i \sigma}^{+}\hat{a}_{j \sigma'}^{+}\hat{a}^{}_{j \sigma'}\hat{a}^{}_{i \sigma} + \frac{1}{2}\sum_{i j,\sigma\sigma'}J^{F}_{ij}\,\hat{a}_{i \sigma}^{+}\hat{a}_{j \sigma'}^{+}\hat{a}^{}_{i \sigma'}\hat{a}^{}_{j \sigma},
\label{Ham}
\end{eqnarray}
where $i(j)$ and $\sigma (\sigma')$ are site and spin indices; $U$, $V_{ij}$ and $J^{F}_{ij}$ represent the local Coulomb, non-local Coulomb and non-local exchange interactions, respectively. $t_{ij}^{\sigma \sigma'}$ is the element of the hopping matrix with spin--orbit coupling.

\begin{table}[!b]
\centering
\caption [Bset]{ Bare non-local exchange interactions and spreads of the Wannier functions calculated for the adatom systems Si(111):\{C,Si,Sn,Pb\}. The third row gives contributions of the adatom $p_{z}$ orbital to the electron density described by the Wannier functions. $M_{S}^{}$ and $M^{\rm adatom}_{S}$ are the total spin magnetization of the $\sqrt{3} \times \sqrt{3}$ unit cell and spin moment of the adatom as obtained from spin-polarized DFT+SO calculations for the ferromagnetic state.}
\label {basisset}
\begin{ruledtabular}
\begin {tabular}{l|cccc}

X & C & Si & Sn & Pb \\
\hline
$J^{F}_{bare}$, meV & 1.64 & 3.81 & 5.44 & 7.34 \\
Spread of WF, \AA$^{2}$ & 12.4 & 15.6 & 16.8 & 17.7 \\ 
$p_{z}$-state in WF, \% & 12 & 42 & 37 & 28 \\ 
\hline

$M_{S}^{}$, $\mu_{B}$ & 0.99 & 0.7 & 0.27 & 0.18 \\
$M^{\rm adatom}_{S}$, $\mu_{B}$ & 0.028 & 0.058 & 0.015 & 0.006 \\
\end {tabular}
\end{ruledtabular}
\label{tab:Bare_parameters}
\end {table}

{\it Coulomb and direct exchange interactions.}
The detailed analysis of the local and non-local Coulomb interactions in the Si(111):X systems was reported in Ref. \onlinecite{Hansmann}. It was found that the screened Coulomb interactions calculated within random phase approximation (RPA) are about 4-5 times smaller than bare ones. In our work we use their partially screened values as reported in Ref.\onlinecite{Hansmann}: $U$ = 1.4, 1.1, 1.0 and 0.9 eV for C, Si, Sn and Pb adatoms, respectively, and $V_{01}$ = 0.5 eV for all adatoms.   

In contrast to previous studies our model contains ferromagnetic exchange interactions as a result of the direct overlap between neighboring Wannier functions. To estimate upper bound of $J^{F}_{ij}$, which corresponds to its bare value, we performed numerical integrations of the following expression by means of Monte Carlo method: 
\begin{equation}
J^{F}_{ij} = \int \frac{W_{i}^{*}(\boldsymbol{r})W_{j}(\boldsymbol{r})W_{j}^{*}(\boldsymbol{r}')W_{i}(\boldsymbol{r}')}{\boldsymbol{r}-\boldsymbol{r}'}  d\boldsymbol{r} d\boldsymbol{r}',
\label{J_F}
\end{equation}
where $W_i(\boldsymbol{r})$ is Wannier function centered on $i$th site.

The results are presented in Table \ref{tab:Bare_parameters}. One can see that the calculated values of $J^{F}_{ij}$ are of millielectronvolt scale and much smaller compared to Coulomb interactions. However, as we will show below, they play an important role in magnetic properties of Si(111):\{C,Si,Sn,Pb\}.

\begin{table}[!h]
\centering
\caption [Bset]{Hopping integrals (in meV) between nearest and next nearest neighbors as obtained from DFT+SO calculations for adatom systems Si(111):\{X\} with X=C, Si, Sn, Pb. See Fig. \ref{im:Crystal}b for details.}
\begin{ruledtabular}
\begin {tabular}{l|cc}
\small
 X  & t$_{01}$ & t$_{02}$ \\
\hline
C &  $\left( \begin{array}{cc} 35.11  &  0.27   \\ -0.27  & 35.11  \end{array} \right)$ &  $\left( \begin{array}{cc} -13.47 + 0.14 i &  -0.43 i \\ -0.43 i & -13.47 - 0.14 i \end{array}\right)$ \\
\\
	Si &  $\left( \begin{array}{cc} 48.33  &  0.71   \\ -0.71  & 48.33  \end{array}\right)$  & $\left( \begin{array}{cc} -20.28 + 0.09 i &  -0.21 i  \\ -0.21 i & -20.28 - 0.09 i \end{array}\right)$ \\
\\
Sn & $\left(\begin{array}{cc} 43.51  &  5.53   \\ -5.53  & 43.51  \end{array}\right)$ &   $\left( \begin{array}{cc} -18.99 + 0.14 i &  -0.86 i  \\ -0.86 i& -18.99 - 0.14 i \end{array}\right)$ \\
\\
Pb &  $\left(\begin{array}{cc} 41.32  &  16.68   \\ -16.68 & 41.32  \end{array} \right)$ &  $\left( \begin{array}{cc} -19.15 + 0.11 i &  -2.09 i \\  -2.09 i & -19.15 - 0.11 i \end{array}\right)$ 
\normalsize
\end {tabular}
\end{ruledtabular}
\label{tab:hoppings}
\end {table}

Due to the smallness of  $J^{F}_{ij}$ direct calculations of its partially screened value within RPA is a hard numerical problem requiring extremely accurate integration. To give a reasonable estimation of the partially screened direct exchange interaction we use the ratio between bare and partially screened values of the Coulomb interaction parameters obtained in Ref.\onlinecite{Hansmann}, that is about 4.5. Thus, one obtains $J^{F}_{01}$ = 0.36, 0.85, 1.21 and 1.63 meV for C, Si, Sn and Pb adatoms, respectively. However, since the determination of $J_{01}^{F}$ is a delicate task, we will also use $J^{F}_{ij}$ as a parameter for description of the ground (Section IV) and excited (Section V) states of Si(111):\{C,Si,Sn,Pb\} by varying its value from zero to its bare limit.

{\it Hopping integrals.}
The calculated hopping integrals are presented in Table \ref{tab:hoppings}. Their diagonal parts are in excellent agreement with previously reported values obtained without spin-orbit coupling.\cite{Hansmann} However, the latter gives a significant contribution that results in comparably large anisotropy.

\section{Hartree-Fock simulations of the electronic models}
\par Computational methods combining first-principles band structure calculations and many-body techniques are of great interest in the physics of strongly correlated materials. Conventional approaches based on density functional theory (DFT) have the well-known difficulties related to a proper treatment of electronic correlations. On the other hand, their extensions taking into account correlations beyond DFT (such as DFT$+U$ and dynamical mean-field theory DMFT) become really involved when a magnetic ground state and spin-orbit coupling effects are concerned. 
\par For example, geometric frustrations and their interplay with electronic correlations have been a subject of intense research. In this context, the class of adatom systems Si(111):X is an ideal candidate to study these effects. It is known that the Hubbard model at half-filling on a triangular lattice displays a 120$^{\circ}$ non-collinear ordering (120$^{\circ}$ N\'eel). However, this point is not verified for the Si(111):Sn system, where early \emph{ab-initio} simulations in the weakly correlated regime showed that the 120$^{\circ}$ N\'eel order is indeed stabilized in the Si(111):Sn system,\cite{Tosatti,Lechermann} while other studies based on the DMFT approach argued that an unusual collinear row-wise (RW) alignment takes place and emerges from long-range electron hopping processes.\cite{Li,Li1} Generally, a geometrically frustrated arrangement may destroy any long-range magnetic configuration and give rise to a spin liquid state. 
\par The problem gets even more complicated when electrons are delocalized. This issue was studied in Ref. \onlinecite{Lechermann} for the Si(111):Sn system, where local magnetic moments residing on Sn adatoms were shown to be small ($\sim 0.06$ $\mu_{\mathrm{B}}$) compared to the total ferromagnetic moment. As it is shown in Table \ref{tab:Bare_parameters}, this takes place in all four systems. Thus, the magnetism in Si(111):X is far from being purely local and has a significant non-local character, so the picture of localized atomic magnetic moments used in DFT calculations seems to be inappropriate.

\begin{figure}[t]
\includegraphics[width=0.49\textwidth]{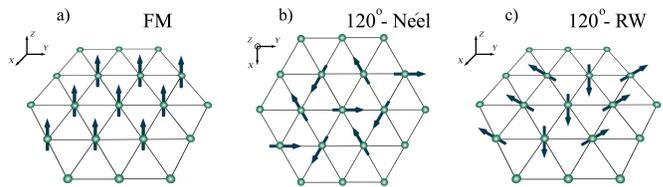}
\caption{Schematic view of magnetic structures used in the Hartree-Fock calculations: a) ferromagnetic (FM), b) 120$^{\circ}$-N\'eel and c) 120$^{\circ}$-RW.}
\label{im:magn}
\end{figure} 

\par To tackle these problems one has to resort to a proper theoretical method. To this end, the basis of Wannier functions seems to be a more appropriate choice compared to that of atomic orbitals, as it incorporates hybridization effects and can serve as an alternative basis for the low-energy model. Indeed, as it is seen in Fig. \ref{im:Wannier}, the resulting Wannier functions constructed by projecting a single band located near the Fermi level onto adatom $p_{z}$ orbitals have a rather complicated structure and are spread in space quite significantly. Nonetheless,  this choice allows us to work in the framework of localized magnetic moments, which in this case reside on the corresponding Wannier function rather than on a single atomic orbital.
\par The effective model Eq. (\ref{Ham}) constructed in the basis of Wannier functions is solved in the mean-field Hartree-Fock approximation, which is proven to be a good tool to study magnetic states in systems with strong correlations:   
\begin{equation}
\left(\hat{t}_{\boldsymbol{k}}+\hat{\mathcal{V}}^{H}_{\boldsymbol{k}} + \hat{\mathcal{J}}^{H}_{\boldsymbol{k}} \right)|\varphi_{\boldsymbol{k}}\rangle=\varepsilon_{\boldsymbol{k}}|\varphi_{\boldsymbol{k}}\rangle,
\label{hf}
\end{equation}
\noindent where $\hat{t}_{\boldsymbol{k}}$ is the Fourier transform of the hopping parameters $\hat{t}_{ij}$ and $\hat{\mathcal{V}}^{H}_{\boldsymbol{k}}$ and $\hat{\mathcal{J}}^{H}_{\boldsymbol{k}}$ are the Hartree-Fock potentials describing the on-site and intersite Coulomb and non-local exchange interactions, respectively, $\varepsilon_{\boldsymbol{k}}$ and $|\varphi_{\boldsymbol{k}}\rangle$ are the corresponding eigenvalues and eigenvectors in a given basis; a self-consistent solution of Eq.~(\ref{hf}) is achieved with respect to the density matrix:
\begin{equation}
\hat{n}=\sum\limits_{\boldsymbol{k}}|\varphi_{\boldsymbol{k}}\rangle \langle\varphi_{\boldsymbol{k}}|.
\end{equation}
\noindent Further details on the computational scheme are provided in Refs. \onlinecite{Solovyev1} and \onlinecite{Nikolaev}.

\begin{figure}[t]
\includegraphics[width=0.48\textwidth]{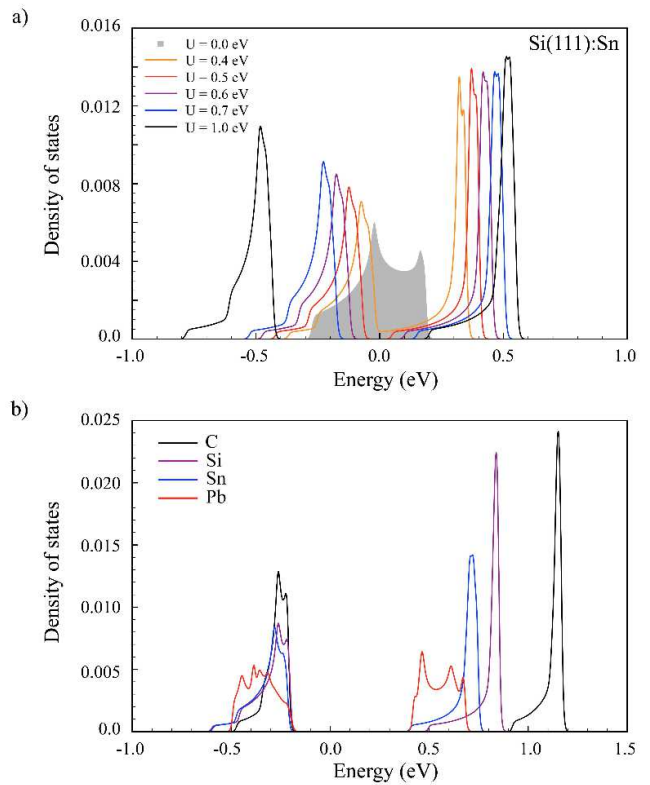}
\caption{a) Densities of states in Si(111):Sn as obtained from the Hartree-Fock approximation for the 120$^{\circ}$-N\'eel magnetic order with different values of $U$ ($V_{01}=0$, $J^{F}_{01}=0$). b) Densities of states corresponding to the magnetic ground states of Si(111):X, X=\{C, Si, Sn, Pb\} as obtained from the Hartree-Fock approximation in the full model, Eq. (\ref{Ham}) (see Section III and Fig. \ref{im:hf}a).} 
\label{im:hf_dos}
\end{figure}

\begin{table}[!b]
\centering
\caption [Bset]{Energy of magnetic configurations (in eV with respect to the ferromagnetic state) in Si(111):X, X=\{C, Si, Sn, Pb\}, as calculated from the Hartree-Fock approximation with $J^{F}_{01}$ = 0 eV.}
\label {basisset}
\begin{ruledtabular}
\begin {tabular}{c|cccc}
  & Si(111):C & Si(111):Si & Si(111):Sn & Si(111):Pb \\
\hline

FM  & 0.0  & 0.0  & 0.0  & 0.0 \\
120$^{\circ}$-N\'eel & $\boldsymbol{-0.055}$   & $\boldsymbol{-0.149}$  & $\boldsymbol{-0.141}$   &  $-0.136$\\
120$^{\circ}$-RW & $-0.042$  & $-0.120$  &  $-0.124$  & $\boldsymbol{-0.143}$  \\
\end {tabular}
\end{ruledtabular}
\label{tab:energyHF}
\end {table}

\begin{figure}[t]
\includegraphics[width=0.48\textwidth]{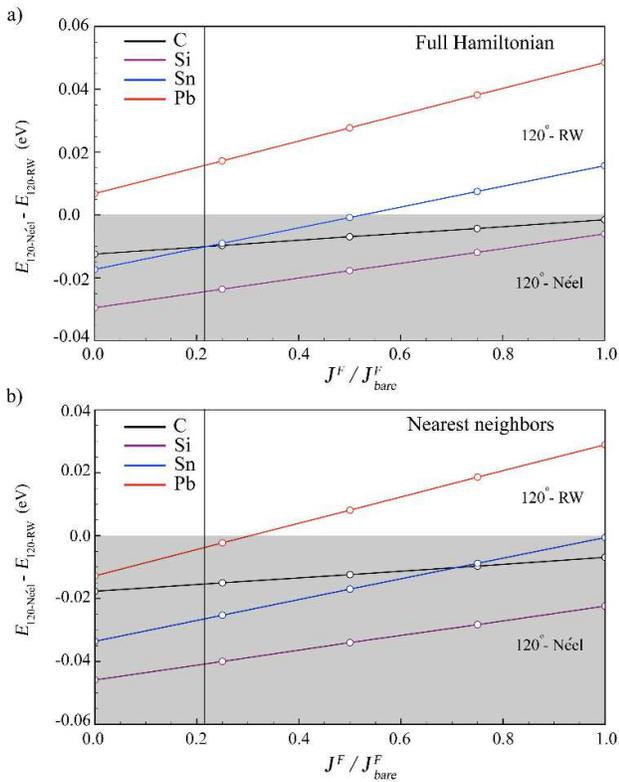}
\caption{Magnetic ground states for different values of $J^{F}$ as obtained from the Hartree-Fock approximation a) in the full model and  b) in the model with nearest neighbor hopping parameters only. The ratio $J_{01}^{F}/J^{F}_{bare}\approx0.22$ given in Section III is shown with vertical lines.}
\label{im:hf}
\end{figure}

\par We have considered three possible magnetic configurations shown in Fig. \ref{im:magn} by comparing their energies calculated within the Hartree-Fock approximation. As a first step, we neglect the non-local exchange interaction $J_{01}^{F}$ and take into account only Coulomb interactions $U$ and $V_{01}$ in Eqs. (\ref{Ham}) and (\ref{hf}). From the corresponding energies presented in Table \ref{tab:energyHF} it is seen that the 120$^{\circ}$-N\'eel order is found to be dominating in Si(111):\{C, Si, Sn\}, while the 120$^{\circ}$-RW order is more favorable in Si(111):Pb. Firstly, it is worth noting that the 120$^{\circ}$-RW magnetic structure is different from the collinear ferrimagnetic order considered in previous studies. Indeed, geometrical frustrations and spin-orbit coupling tend to align magnetic moments to form a 120$^{\circ}$ structure in the $xz$ plane. Secondly, the 120$^{\circ}$-N\'eel order in X=Sn is in agreement with previous studies based on DFT+$U$ calculations, while the DMFT based approaches predict RW to be a magnetic ground state for sufficiently large values of $U$. First of all, this discrepancy can be attributed to the fact that the Hartree-Fock approximation is formulated at zero temperature, while previous studies based on DMFT have been performed in the experimentally accessible temperature range. Next, our model Eq. (\ref{Ham}) is extended to include the effect of spin-orbit coupling, which in the case of Si(111):\{Sn, Pb\} gives a significant contribution renormalizing hopping parameters.
\par To give some comparison on different approaches, we have explored critical values of the on-site Coulomb interaction $U$ in a metal-insulator phase transition. Density of states of the Si(111):Sn system calculated within the Hartree-Fock approximation for different values of $U$ ($V_{01}=0$, $J_{01}^{F}=0$) are shown in Fig.\ref{im:hf_dos}a. As it is seen, the charge gap starts to open at $U_{c}\approx 0.5$ eV, that is smaller than the critical values $U_{c}\approx$  0.60, 0.65 and 0.75 eV obtained within the single-site DMFT, variational cluster and dual fermion approaches, respectively.\cite{Lechermann,Li1} Despite this fact, we believe that the Hartree-Fock approximation is still reliable since the values of $U$ used in our calculations (see Section III) are much higher compared to the critical ones. However, even though we treat electronic correlations in a mean-field manner, they do play an important role in stabilizing a magnetic ground state in the Si(111):X system.
\par Next, we proceed to study the effect of the non-local exchange interaction $J_{01}^{F}$ on a magnetic ground state. The results obtained with respect to the ratio $J_{01}^{F}/J^{F}_{bare}$ are presented in Fig. (\ref{im:hf})a. One can see that for any value of $J_{01}^{F}/J^{F}_{bare}$ the 120$^{\circ}$-N\'eel order is found to be stable in X=C and Si, while the 120$^{\circ}$-RW magnetic structure is stabilized only in X=Pb. The situation is different in the case of Si(111):Sn, where depending on the ratio of $J_{01}^{F}/J^{F}_{bare}$ both magnetic states can be realized. This result leads to a very interesting conclusion that the magnetic ground state in the Si(111):X systems is also controlled by the value of $J^{F}$.
\par To get a deeper insight on this effect, we eliminate hopping parameters beyond nearest neighbors and perform the same calculations with respect to the ratio of $J_{01}^{F}/J^{F}_{bare}$. As it is shown in Fig. \ref{im:hf}b, the 120$^{\circ}$-N\'eel order is stabilized in all four systems, as expected for the nearest-neighbor Hubbard model on a triangular lattice. However, in this case the transition between the 120$^{\circ}$ N\'eel and 120$^{\circ}$-RW magnetic orders is still observed in Si(111):Pb. As will be shown below, this non-local exchange interaction gives an additional contribution to the kinetic isotropic exchange parameters between magnetic moments favoring their ferromagnetic alignment. Meanwhile, we conclude that the stabilization of a magnetic order in the Si(111):X systems is a joint effect of long-range hopping processes, spin-orbit coupling and non-local electron correlations.

\section{Spin Hamiltonian}
To probe excited magnetic states in the adatom systems we construct spin models within the superexchange theory\cite{Anderson} formulated in the limit $t_{ij}\ll U$. In our case $t_{ij}/U$ varies from 0.025 for Si(111):C to 0.045 for Si(111):Pb that justifies this approach. The corresponding spin Hamiltonian is given by:
\begin{eqnarray}
\hat {\mathcal{H}}^{spin} = \sum_{ij} J_{ij} \hat {\boldsymbol{S}}_{i} \hat {\boldsymbol{S}}_{j} + \sum_{ij} \boldsymbol{D}_{ij}  [\hat {\boldsymbol{S}}_{i} \times  \hat {\boldsymbol{S}}_{j}] + \sum\limits_{ij}\hat{\mathbf{S}}_i\overset{\leftrightarrow}{\Gamma}_{ij}\hat{\mathbf{S}}_j, 
\label{spinham}
\end{eqnarray}
where $\hat {\boldsymbol{S}}$ is the spin operator. $J_{ij}$, $\boldsymbol{D}_{ij}$ and $\overset{\leftrightarrow}{\Gamma}_{ij}$ are the isotropic exchange coupling, antisymmetric anisotropic (Dzyaloshinskii-Moriya) and symmetric anisotropic interactions, respectively. The summation runs twice over all pairs.

{\it Isotropic exchange interaction.}
In terms of the electronic model parameters given in Eq.~(\ref{Ham}) the isotropic exchange interaction can be expressed in the following form:\cite{Anderson, Aharony}
\begin{eqnarray}
J_{ij} = \frac{1}{\widetilde U} {\rm Tr_{\sigma}} \{ \hat t_{ji} \hat t_{ij} \} -  J^{F}_{ij},
\label{exch}
\end{eqnarray}
where $\hat t_{ij}$ is the hopping integral with spin-orbit coupling, the effective local Coulomb interaction is estimated as $\widetilde U = U - V_{ij}$. The first kinetic term is the famous Anderson's superexchange. In turn, the second one, $J_{ij}^{F}$ represents the direct ferromagnetic exchange due to the overlap between neighboring Wannier functions. Table \ref{tab:DM_vectors} gives the values of the isotropic interactions calculated with the partially screened $J_{ij}^{F}$ as described in Section III. 

{\it Anisotropic exchange interactions,} antisymmetric Dzyaloshinskii-Moriya and symmetric anisotropic exchange interactions are given by
\begin{equation}
\mathbf{D}_{ij} = \frac{i}{2\widetilde{U}}[{\rm Tr}(\hat {t}_{ij}){\rm Tr}(\hat{t}_{ji}\boldsymbol{\sigma})-{\rm Tr}(\hat{t}_{ji}){\rm Tr}(\hat{t}_{ij}\boldsymbol{\sigma})],
\label{eq:DM-vector}
\end{equation}
\begin{equation}
\overset{\leftrightarrow}{\Gamma}_{ij} = \frac{1}{2\widetilde{U}}[{\rm Tr}(\hat{t}_{ji}\boldsymbol{\sigma})\otimes {\rm Tr}(\hat{t}_{ij}\boldsymbol{\sigma})+{\rm Tr}(\hat{t}_{ij}\boldsymbol{\sigma})\otimes {\rm Tr}(\hat{t}_{ji}\boldsymbol{\sigma})],
\label{eq:Gamma-matrix}
\end{equation}
where $\boldsymbol{\sigma}$ are the Pauli matrices. 

The calculated DMIs are presented in Table \ref{tab:DM_vectors}. Let us first discuss their symmetry. Since the resulting Wannier functions reside on the adatom-silicon bonds, symmetry properties of the spin Hamiltonian are consistent with the  $C_{3v}$ point group of the triangular lattice formed by adatoms. According to Moriya's rules \cite{Moriya}, vertical reflections go through the bonds between nearest neighbours, and the corresponding anisotropic exchange parameters are perpendicular to their bonds and lie in the $xy$ plane. On the other hand, next-nearest neighbours are not located on the mirror planes, and we obtain the non-zero $z$ components of the anisotropic exchange parameters that alternate within the coordination sphere.

In the systems with inversion symmetry breaking the ratio $\frac{|\mathbf{D}_{ij}|}{J_{ij}}$ is a control parameter for the period of spiral structures or size of the individual skyrmion at finite temperatures and magnetic fields. Depending on the adatom this ratio for the kinetic interactions presented in Table  \ref{tab:DM_vectors} is varied from 0.017 (for X=C) to 0.83 (for X=Pb). It provides unprecedented possibilities to control and tune the DMI strength within this family of surface nanostructures.

Another important contribution to the magnetic anisotropy is the symmetric anisotropic exchange interaction, $\overset{\leftrightarrow}{\Gamma}_{ij}$.  
The calculated tensors for the Si(111): Sn and Pb systems are presented in Table \ref{tab:Gamma}. One can see that they favor $xz$ plane alignment of the magnetic moments. Thus, the principal axis of $\overset{\leftrightarrow}{\Gamma}_{01}$ coincides with the direction of DMI for the same bond. It agrees with the results of Ref.\onlinecite{Aharony1} where a general one-band Hubbard model with spin-orbit coupling was analyzed. We also found that the elements of $\Gamma_{ij}$ for X=Si and C are less than 10$^{-4}$ meV.

Importantly, the spin Hamiltonians obtained for the adatom systems can be classified with respect to the ratio between nearest-neighbor DMI, $\mathbf{D}_{01}$ and next-nearest neighbor isotropic exchange interaction, $J_{02}$. For instance, in the case of the Si(111):C and Si(111):Si systems $J_{02} > |\mathbf{D}_{01}|$ and, therefore, the spin model is the isotropic one of $J_{1}-J_{2}$ type. The ratio $\frac{J_{1}}{J_{2}}$ is close to 10, which prevents the formation of an incommensurate spiral structure in the ground state as well as a skyrmion state at finite magnetic fields. The critical ratio $\frac{J_{1}}{J_{2}}$ can be substantially decreased up to 5, if one takes the bare value of the direct exchange interaction between nearest neighbours in Eq.(\ref{exch}). Nevertheless, this ratio is also beyond the limit $\frac{J_{1}}{J_{2}} < 1$ favoring the formation of skyrmions.\cite{Okubo}  

The situation is different in the case of Sn/Si(111) and Pb/Si(111), for which nearest neighbor $J_{01}$ and $\mathbf{D}_{01}$ are of the same order and much larger than $J_{02}$. Namely, this property, as we will show in the next section, leads to the antiferromagnetic skyrmion lattice state.

\begin{table}[!h]
\centering
\caption [Bset]{Isotropic $J_{ij}$ and anisotropic $\boldsymbol{D}_{ij}$ exchange interactions (in meV) in Si(111):X, X=\{C, Si, Sn, Pb\} as obtained from DFT+SO calculations, Eqs. (\ref{exch}) and (\ref{eq:DM-vector}). See Fig. \ref{im:Crystal}b for details.}
\label {basisset}
\begin{ruledtabular}
\begin {tabular}{l|cccc}
 X  & $J_{01}$ & $\boldsymbol{D}_{01}$ & $J_{02}$ &  $\boldsymbol{D}_{02}$  \\
 \hline
C & 2.38 &  (0.0, 0.042, 0.0) & 0.26 & (0.015, 0.0, -0.005)\\
\\
Si & 6.94 & (0.0, 0.228, 0.0) & 0.75 & (0.015, 0.0, -0.005)\\
\\
Sn & 6.48 & (0.0, 1.925, 0.0) & 0.73 & (0.065, 0.0, -0.010) \\
\\
Pb & 8.30 & (0.0, 6.895, 0.0) & 0.83 & (0.180, 0.0, -0.009)\\
\end {tabular}
\end{ruledtabular}
\label{tab:DM_vectors}
\end {table}

\begin{table}
\centering
\caption [Bset]{Symmetric anisotropic exchange interactions $\overset{\leftrightarrow}{\Gamma}_{01}$  (in meV) in Si(111):Sn and Si(111):Pb as obtained from DFT+SO calculations, Eq. (\ref{eq:Gamma-matrix}). See Fig. \ref{im:Crystal}b for details.}
\label {basisset}
\begin{ruledtabular}
\begin {tabular}{l|cc}
 X & Sn & Pb \\
 \hline
$\overset{\leftrightarrow}{\Gamma}_{01}$ & $\left( \begin{array}{ccc} 0.0  &  0.0 & 0.0  \\ 0.0  & 0.245 & 0.0 \\ 0.0 & 0.0 & 0.0  \end{array}\right)$ & $\left( \begin{array}{ccc} 0.0  &  0.0 & 0.0  \\ 0.0  & 2.784 & 0.0 \\ 0.0 & 0.0 & 0.0  \end{array}\right)$ \\
\end {tabular}
\end{ruledtabular}
\label{tab:Gamma}
\end {table}

\begin{figure*}[t]
\includegraphics[width=0.98\textwidth]{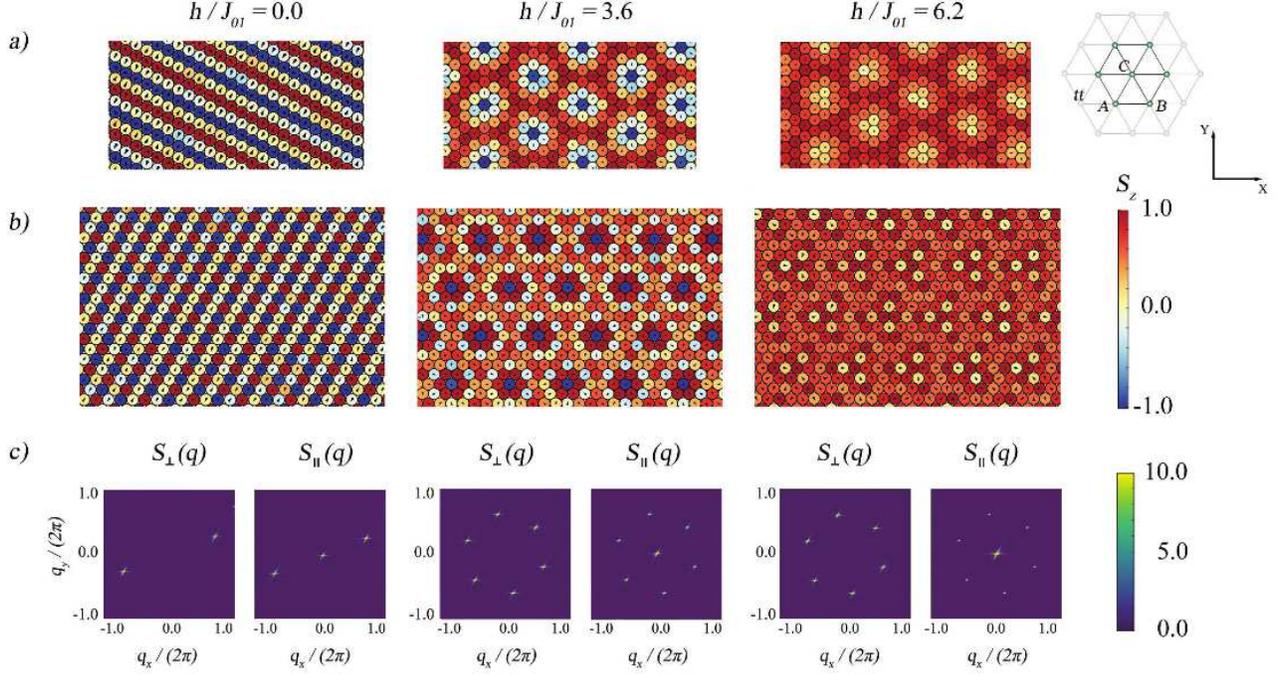}
\caption{Snapshots of the Si(111):Pb spin texture given for a sublattice $A$ (a) and a full lattice (b) as obtained from Monte Carlo simulations for $N=150\times150$, $T/J_{01}=0.01$ and different values of $h/J_{01}$. Spin components in the $xy$ plane are indicated with black arrows. c) Static spin structure factors for the corresponding spin textures. } 
\label{im:skyrm}
\end{figure*}

\section{Monte Carlo results}
In crystals with the $C_{nv}$ symmetry, the anisotropic exchange interaction favors a rotation of magnetic moments along the propagation direction of a spin spiral structure, and they are expected to  possess a N\'eel-type skyrmion state.\cite{Bogdanov1} Moreover, the formation of an antiferromagnetic skyrmion texture (AF-SkX) on the antiferromagnetic triangular lattice with Dzyaloshinskii-Moriya interactions was reported recently.\cite{Rosales}
\par In this section we focus on the effect of an external magnetic field $\boldsymbol{h}$ applied to the spin system Eq. (\ref{spinham}):
\begin{equation}
\mathcal{H}=\mathcal{H}^{spin}-\boldsymbol{h}\cdot\sum_{i}\boldsymbol{e}_{i},
\end{equation}
\noindent where the spin variables are now treated as classical vectors, $|\boldsymbol{e}_{i}|=1$. In a classical limit for the given spin vector length one has to renormalize model parameters of the quantum spin model. This is done by scaling the exchange interactions (given in Table IV and V) by the maximum length of the product of two spin operators, that is $S(S+1)$, where $S=1/2$ and $\hbar=1$. However, it is worth noting that this scaling is rather arbitrary and instead of using unit vectors one can leave their quantum mechanical length without distinction between model parameters. 
\par Our Monte Carlo simulations have been performed based on the heat-bath method combined with overrelaxation. The corresponding model parameters are given up to next-nearest neighbors. In these calculations supercells of various size from $N=96\times96$ to $150\times150$ spins with periodic boundary conditions are used and a single run contains $(0.5-2.0)\cdot 10^{6}$ Monte Carlo steps. For initial relaxation the system is gradually cooled down from higher temperatures. 

\par While different states can be identified from the real-space spin textures, to trace their formation we have computed the static spin structure factors:
\begin{equation}
S_{\perp}(\boldsymbol{q})=\frac{1}{N}\left\langle\left|\sum_{i} e_{i}^{x} \, e^{-i\boldsymbol{q}\cdot\boldsymbol{r}_{i}} \right|^{2}+\left|\sum_{i} e_{i}^{y} \, e^{-i\boldsymbol{q}\cdot\boldsymbol{r}_{i}} \right|^{2}\right\rangle
\end{equation}
\noindent and
\begin{equation}
S_{\parallel}(\boldsymbol{q})=\frac{1}{N}\left\langle\left|\sum_{i} e_{i}^{z} e^{-i\boldsymbol{q}\cdot\boldsymbol{r}_{i}} \right|^{2}\right\rangle,
\end{equation}
\noindent as well as the total chirality $\chi_{L}$ and skyrmon number $\chi_{Q}$:
\begin{equation}
\chi_{L}=\frac{1}{8\pi}\left\langle\sum_{i}\chi_{i}^{(12)}+\chi_{i}^{(34)} \right\rangle
\end{equation}
\noindent and
\begin{equation}
\chi_{Q}=\frac{1}{8\pi}\left\langle\sum_{i}A_{i}^{(12)}\mathrm{sgn}[\chi_{i}^{(12)}]+A_{i}^{(34)}\mathrm{sgn}[\chi_{i}^{(34)}] \right\rangle,
\end{equation}
\noindent where $\chi_{i}^{(ab)}=\boldsymbol{e}_{i}\cdot\boldsymbol{e}_{a}\times\boldsymbol{e}_{b}$ is the so-called local chirality defined on a triangle $\{\boldsymbol{r}_{i},\boldsymbol{r}_{a},\boldsymbol{r}_{b}\}$ and $A_{i}^{(ab)}=\parallel(\boldsymbol{e}_{a}-\boldsymbol{e}_{i})\times(\boldsymbol{e}_{b}-\boldsymbol{e}_{i}) \parallel/\,2$ is the corresponding area. The latter quantities are considered as order parameters that represent topological stability of the corresponding state. 
\par
\begin{figure}[t]
\includegraphics[width=0.49\textwidth]{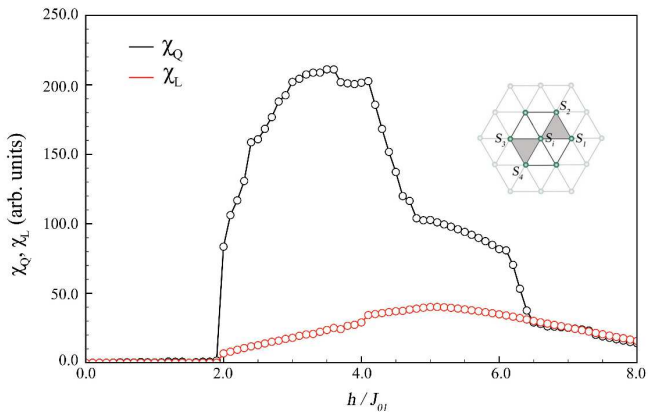}
\caption{Skyrmion number $\chi_{Q}$ and total chirality $\chi_{L}$ as a function of the applied magnetic field $h/J_{01}$ obtained from Monte Carlo simulations for $N=150\times150$ and $T/J_{01}=0.01$ in Si(111):Pb. The inset shows the corresponding area used to calculate local chiralities.}
\label{im:chi}
\end{figure}

 The results obtained for the Si(111):Pb system are given in Fig. \ref{im:skyrm}. Interestingly, the system exhibits several phases as a magnetic field applied along the $z$ axis is varied. At low magnetic fields a complex spin spiral state is stabilized. It is comprised of three interpenetrating spin spirals formed on each sublattice and characterized by a single $\boldsymbol{q}$-vector. As the magnetic field increases the system enters into a stable AF-SkX state which is a superposition of three N\'eel type SkX lattices characterized by three $q$ vectors (which are in turn formed by three spin spirals). As it is seen from Fig. \ref{im:chi}, the AF-SkX state is favored in a wide range of magnetic fields. However, a stepwise behavior of the skyrmion number and total chirality with respect to the magnetic field is a result of the discrete finite-size model allowing for only definite numbers of skyrmions.\cite{Rosales} Finally, at higher magnetic fields the AF-SkX state is followed by a vortex-like texture and a paramagnetic phase. Our results are in agreement with those reported in Ref.\onlinecite{Rosales}. However, it is worth mentioning that in this work we have employed an extended spin model including both antisymmetric and symmetric anisotropy terms up to the next-nearest neighbors, that justifies the realization of the so-called multiple $\boldsymbol{q}$-states in a more general case. 

Similar skyrmion lattice state is realized in Sn/Si(111). However, there is one important difference, that is a skyrmion size controlled by the ratio $\frac{|\mathbf{D}_{ij}|}{J_{ij}}$. In the case of Si(111):Sn and Si(111):Pb the skyrmion size is estimated to be about 40 \AA \, and 26 \AA, respectively.

Fig. \ref{im:chi} gives the value of critical fields needed to form a skyrmion state, that is $\sim 2J_{01}$. Taking the estimated g-factor and calculated exchange interactions we conclude that the critical point may be accessible in magnetic fields $\sim 190$ T for Sn/Si(111) and $\sim 250$ T for Pb/Si(111). These fields are too large to be reached in laboratories. To decrease their values one has to reduce isotropic exchange interactions between nearest neighbours.  In our simulations it can be done by changing the value of $J^{F}_{01}$. For instance, if one takes its bare value $J^{F}_{bare}=5.44$ meV for Sn/Si(111) the critical field can be estimated as 66 T. On the other hand one can consider a mixed adatom system combining carbon (weak isotropic exchange) and tin (strong Dzyaloshinksii-Moriya interaction) sublattices. This aspect remains open for future investigation.

\section{Summary}
The main purpose of our study is to complete the picture of principal interactions in the Si(111):\{C,Si,Sn,Pb\} adatom systems. Taking into account spin-orbit coupling leads to a complex non-diagonal form of the hopping matrix, while the overlap between neighboring Wannier functions is responsible for the direct ferromagnetic exchange interaction that strongly affects low-energy properties of the systems in question. Our solutions of the constructed electronic and spin models have shown that the resulting state of the surface nanosystem mainly depends on these parameters that can be varied with the adatom type and their coupling with a substrate.

\emph{Acknowledgment}. We acknowledge fruitful discussions with Igor Solovyev. The work is supported by the Ministry of Education and Science of the Russian Federation, Project No. 16.1751.2014/K and the grant of the President of Russian Federation MD-6458.2016.2.


\begin{thebibliography}{00}

\bibitem{Modesti}
S. Modesti, L. Petaccia,  G. Ceballos, I. Vobornik, G. Panaccione, G. Rossi, L. Ottaviano, R. Larciprete, S. Lizzit, and A. Goldoni, Phys. Rev. Lett. 98, 126401 (2007). 

\bibitem{Tosatti}
G. Profeta   and E. Tosatti, Phys. Rev. Lett. 98, 086401  (2007). 

\bibitem{Carpinelli}
J. M. Carpinelli, H. H. Weitering, M. Bartkowiak, R. Stumpf, and E. W. Plummer, Phys. Rev. Lett. 79, 2859 (1997). 

\bibitem{Slezak}
J. Slez\'ak, P. Mutombo, V. Ch\'ab,
Phys. Rev. B 60, 13328 (1999).

\bibitem{Hansmann}
P. Hansmann, T. Ayral, L. Vaugier, P. Werner, and S. Biermann, Phys. Rev. Lett. 110, 166401  (2013).

\bibitem{Hansmann_1}
P. Hansmann, L. Vaugier, H. Jiang, S. Biermann, Journal of Physics: Condensed Matter 25, 094005 (2013). 

\bibitem{Li}
G. Li, P. Hopfner, J. Schafer, C. Blumenstein, S. Meyer, A. Bostwick, E. Rotenberg, R. Claessen and W. Hanke , Nature Communications  4, 1620  (2013).

\bibitem{Li1}
G. Li, M. Laubach, A. Fleszar, and W. Hanke,
Phys. Rev. B 83, 041104 (2011).

\bibitem{Lechermann}
S. Schuwalow, D. Grieger, and F. Lechermann, Phys. Rev. B 82, 035116 (2010).

\bibitem{Hansmann_2}
P. Hansmann, T. Ayral, A. Tejeda and S. Biermann, 
Nature Scientific Reports 6, 19728 (2016).

\bibitem{Fu}
Huixia Fu, Zheng Liu, Chao Lian, Jin Zhang, Hui Li, Jia-Tao Sun, Sheng Meng,
arXiv:1606.08945.

\bibitem{VESTA}
Momma Koichi and Fujio Izumi, J. Appl. Crystallogr. 44, 1272 (2011).

\bibitem{Wiesendanger}
R. Wiesendanger, Rev. Mod. Phys. 81, 1495 (2009).

\bibitem{Wiesendanger1}
N. Romming, A. Kubetzka, C. Hanneken, K. von Bergmann, and R. Wiesendanger,
Phys. Rev. Lett. 114, 177203 (2015).

\bibitem{DFT}
W. Kohn and L. J. Sham,  Phys. Rev. 11, A1133 (1965).

\bibitem{PBE}
John P. Perdew, Kieron Burke, and Matthias Ernzerhof, Phys. Rev. Lett. 77, 3865 (1996).


\bibitem{espresso}
P. Giannozzi, S. Baroni, N. Bonini, M. Calandra, R. Car, C. Cavazzoni, D. Ceresoli, G. L. Chiarotti, M. Cococcioni, I. Dabo, A. Dal Corso, S. de Gironcoli, S. Fabris, G. Fratesi, R. Gebauer, U. Gerstmann, C. Gougoussis, A. Kokalj, M. Lazzeri, L. Martin-Samos, N. Marzari, F. Mauri, R. Mazzarello, S. Paolini, A. Pasquarello, L. Paulatto, C. Sbraccia, S. Scandolo, G. Sclauzero, A. P. Seitsonen, A. Smogunov, P. Umari, and R. M Wentzcovitch,  J. Phys.: Condens. Matter 21, 395502 (2009).


\bibitem{Kresse}
G. Kresse and J. Hafner, 
Phys. Rev. B 47, 558 (1993).

\bibitem{Furthmuller}
G. Kresse and J. Furthm\"uller, 
Phys. Rev. B 54, 11169 (1996).

\bibitem{Pignedoli}
C. A. Pignedoli,  A. Catellani,  P. Castrucci,  A. Sgarlata,  M. Scarselli,  M. De Crescenzi,  and C. M. Bertoni, Phys. Rev. B 69, 113313 (2004).

\bibitem{wannier90}
N. Marzari and D. Vanderbilt, Phys. Rev. B 56, 12847 (1997).

\bibitem{wannier901}
I. Souza, N. Marzari, and D. Vanderbilt, Phys. Rev. B 65, 035109 (2001).

\bibitem{wannier902}
A. A. Mostofi, J. R. Yates, Y-S. Lee, I. Souza, D. Vanderbilt, and N. Marzari, Comput. Phys. Commun. 178, 685 (2008).

\bibitem{Danis}
D. I. Badrtdinov, O. S. Volkova, A. A. Tsirlin, I. V. Solovyev, A. N. Vasiliev, and V. V. Mazurenko,
Phys. Rev. B 94, 054435 (2016).

\bibitem{Vanderbilt}
T. Thonhauser, D. Ceresoli, D. Vanderbilt and R. Resta, Phys. Rev. Lett. 95, 137205 (2005).

\bibitem{Nikolaev_orb}
S.A. Nikolaev and I.V. Solovyev, Phys. Rev. B 89, 064428 (2014).

\bibitem{Solovyev1}
I.V. Solovyev, J. Phys.: Condens. Matter 20, 293201 (2008).


\bibitem{Nikolaev}
S.~A. Nikolaev, V.~V. Mazurenko, A.~A. Tsirlin, and V.~G. Mazurenko, arXiv:1603.05192.

\bibitem{Anderson}
P.W. Anderson, Phys. Rev. 2, 115 (1959).

\bibitem{Aharony}
T. Yildirim, A. B. Harris, Amnon Aharony, and O. Entin-Wohlman, Phys. Rev. B 52, 10239 (1995).


\bibitem{Moriya}
T. Moriya, Phys. Rev. 120, 91 (1960).

\bibitem{Aharony1}
L. Shekhtman, O. Entin-Wohlman and A. Aharony,
Phys. Rev. Lett. 69, 836 (1992).

\bibitem{Okubo}
T. Okubo, S. Chung, and H. Kawamura, Phys. Rev. Lett. {\bf 108}, 017206 (2012).


\bibitem{Bogdanov1}
A. N. Bogdanov and D. A. Yablonsky, Sov. Phys. JETP 95, 178 (1989). 

\bibitem{Rosales}
H.~D. Rosales, D.~C. Cabra, and P. Pujol, Phys. Rev. B {\bf 92}, 214439 (2015).

\end{thebibliography}
\end{document}